\documentclass[10pt,twoside,BCOR7mm,DIV15,headinclude,footexclude,cleardoubleempty,idxtotoc]{scrartcl}

\usepackage[english]{babel}
\usepackage{graphicx}
\usepackage{hyperref}
\usepackage{scrpage2}
\usepackage{hyperref}
\usepackage{ifthen}

\makeatletter
\renewcommand{\@biblabel}[1]{}
\renewcommand{\@cite}[2]{%
{#1\ifthenelse{\boolean{@tempswa}}{,#2}{}}}
\makeatother

\hypersetup{breaklinks=true
,colorlinks=true,linkcolor=black,urlcolor=blue
,citecolor=black}

\ofoot{\thepage}
\ifoot{}

\setheadsepline{1pt}

\setkomafont{pagehead}{\normalfont\sffamily}
\setkomafont{pagenumber}{\normalfont\rmfamily}

\usepackage{booktabs}
\usepackage{amsmath}
\usepackage{amssymb}
\usepackage{multicol}
\usepackage{float}

\makeatletter
\newcommand{\listofcontributions}{\@starttoc{con}}

\newcommand{\l@contribution} {\@dottedtocline{1}{1.5em}{2.3em}}
\makeatother

\newenvironment{contribution}{
\setcounter{section}{0}
\setcounter{figure}{0}
\setcounter{table}{0}
\begin{flushleft}
{\em Clumping in Hot Star Winds \\
W.-R.\ Hamann, A.\ Feldmeier \& L.\ Oskinova, eds.\\
Potsdam: Univ.-Verl., 2007 \\
URN: http://nbn-resolving.de/urn:nbn:de:kobv:517-opus-13981
} 
\end{flushleft}
}{
\newpage
\lehead{}
\rohead{}
}

\begin{document}

\setlength{\baselineskip}{2.5ex}

\begin{contribution}
% PROCEEDINGS OF THE CLUMPING WORKSHOP.
% NOTE THAT YOU MUST NOT PROCESS THIS FILE, BUT THE MASTER FILE:
% latex masterfile; dvips masterfile

% RUNNING AUTHOR: PUT AUTHOR NAMED HERE
\lehead{Jorick S.\ Vink et al.}

% RUNNING TITLE; SHORTEN THE TITLE IF NECESSARY
% IN CASE OF A ONE-PAGE CONTRIBUTION (POSTER),
% SQUEEZE AUTHORS AND TITLE IN THIS LINE (Author: Title ...)
\rohead{Advances in mass-loss predictions}

\begin{center}
% FULL TITLE HEADING
{\LARGE \bf Advances in mass-loss predictions}\\
\medskip

% AUTHORS LIST
{\it\bf Jorick S. \ Vink$^1$, P. \ Benaglia$^2$, B.\ Davies$^3$, A. \ de Koter$^4$ \& R.D. \ Oudmaijer$^5$}\\

% AFFILIATIONS
{\it $^1$Armagh Observatory, Northern Ireland, UK}\\
{\it $^2$Instituto Argentino de Radioastronomia, Argentina}\\
{\it $^3$Rochester Institute of Technology, USA}\\
{\it $^4$Astronomical Institute Anton Pannekoek, University of Amsterdam, The Netherlands}\\
{\it $^5$School of Physics \& Astronomy, University of Leeds, UK}

% ABSTRACT
\begin{abstract}
We present the results of Monte Carlo mass-loss predictions for massive stars
covering a wide range of stellar parameters. We critically test our predictions against 
a range of observed mass-loss rates -- in light of the recent discussions on wind 
clumping. We also present a model to compute the clumping-induced polarimetric 
variability of hot stars and we compare this with observations of Luminous Blue Variables, for 
which polarimetric variability is larger than for O and Wolf-Rayet stars. 
Luminous Blue Variables comprise an ideal testbed for studies of wind clumping and wind geometry, 
as well as for wind strength calculations, and we propose they may be direct supernova progenitors.
\end{abstract}
\end{center}

% TEXT OF THE PAPER, TWO-COLUMN STYLE
\begin{multicols}{2}

\section{Introduction}

This contribution consists of two complementary aspects of hot-star winds.
We first describe the results of mass-loss predictions -- 
widely used in current massive star models in the galaxy, and beyond. 
In particular, we test our predictions as a function of effective temperature against  
recent radio data, and we discuss the potential implications for 
the clumping properties of supergiants of various spectral types (Sect.\ref{vink:mp}). 
We also discuss mass-loss predictions for the winds of Luminous Blue 
Variables (LBVs), and we present results of the clumping-induced 
polarimetric variability of hot-star winds (Sect.~\ref{vink:polvar}), before we conclude.
 
\section{Monte Carlo mass-loss predictions}
\label{vink:mp}

Our method to predict the mass-loss rates of massive stars is based 
on Monte Carlo radiative transfer calculations (see Vink et al. \cite{vink:vink99}, de Koter et al. \cite{vink:dekoter97}). 
In short, we compute non-LTE level populations for all relevant ions from hydrogen to zinc, before 
we follow the fate of a large number of photon packets through the wind. We predict the efficiency of the 
momentum transfer from the photons to the gas, generally assuming a pre-described velocity law 
(but see Vink et al. \cite{vink:vink99}, M\"uller \& Vink, {\it in prep.}). 
We derive wind efficiencies, $\eta$ $=$ $(\dot{M} v_{\infty})/(L/c)$, 
for a range of stellar parameters (including metallicities).
  
\subsection{Results: successes}

To gauge the success of our models and to be able to identify discrepancies, we test our 
predictions against a survey of radio mass-loss rates (Benaglia et al. \cite{vink:benaglia07}) 
from the free-free emission in hot-star winds. 

\begin{figure}[H]
\begin{center}
\includegraphics[width=\columnwidth]{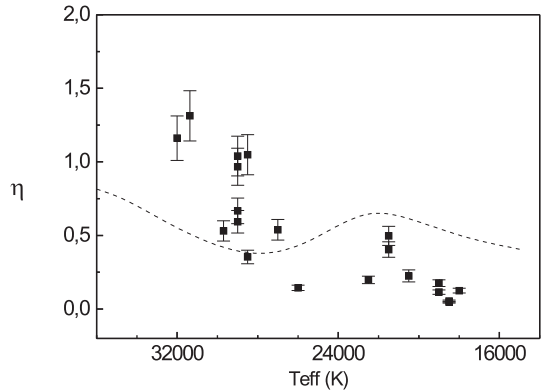}
\caption{Radio wind efficiency vs. effective temperature (Benaglia et al. \cite{vink:benaglia07}). 
Over-plotted
are mass-loss predictions (Vink et al. \cite{vink:vink99}). 
Note the possible bi-stability jump at 21~000 K.}
\label{vink:eta} 
\end{center}
\end{figure}

In Fig.~\ref{vink:eta}, we plot the wind efficiency versus effective temperature 
in the regime of the wind bi-stability, where winds are predicted to change from 
lower $\dot{M}$, fast winds, on the hot side, to higher $\dot{M}$, slow winds on the cool side -- a result 
of a change in the Fe ionisation that drives the wind.  We over-plot the mass-loss predictions 
around the bi-stability jump (dashed line) and focus on the general trends, before we 
continue our discussion on quantitative aspects, and their implications, in Sect~\ref{vink:discr}. 

The overall behaviour shows that $\eta$ declines when the temperature drops. 
At the highest temperatures, the flux and the opacity show large overlap and 
the momentum transfer efficiency is maximal. At lower effective temperature, the flux is 
gradually emitted towards lower wavelengths, and there is a growing mismatch between the 
flux and the ultraviolet line opacity.
Figure~\ref{vink:eta} shows that our predictions of this overall behaviour are confirmed by the radio survey.
Our second prediction is that there should be an increase in the mass-loss rate due to the opacity increase 
when Fe recombines from Fe {\sc iv} to Fe {\sc iii}. Our radio data appear to confirm the presence of a local maximum 
around 21~000 K, although data around this critical temperate is, as yet, sparse. 

The bi-stability limit is relevant for stellar evolution calculations when stars evolve off the main sequence 
towards the red part of the Hertzsprung-Russell diagram (HRD). 
It may also play a role for LBV winds, when LBVs such as AG~Car, change their temperatures -- and radii -- on 
timescales of the S~Dor variations (of the order of years to decades). 
This variable wind behaviour -- predicted by Vink \& de Koter (\cite{vink:vink02}) -- is anticipated to result 
in circumstellar media consisting of concentric shells with varying wind densities.
Kotak \& Vink (\cite{vink:kotak06}) recently suggested that the quasi-periodic modulations seen in the radio lightcurves of 
some supernovae may imply that LBVs could be {\it direct} supernova progenitors. 
At first this seems at odds with stellar evolution calculations, which do not predict
LBVs to explode. However, there is a growing body of evidence hinting that LBVs may nonetheless 
explode (Pastorello et al. \cite{vink:pastorello07}, Smith \cite{vink:smith07}, Gal-Yam et al. \cite{vink:galyam07}).

Despite the success of our models in explaining LBV mass-loss variability, the bi-stability jump, and the 
scaling of $\dot{M}$ with metallicity (see de Koter, this volume), we turn to discrepancies of 
our models against empirical mass-loss rates.

\subsection{Results: discrepancies}
\label{vink:discr}

Discrepancies have been noted between the Vink et al. (\cite{vink:vink00}) predictions and 
empirical mass-loss rates in several areas of the HRD. One group of objects 
is that of low $L$ (log $L/L_{\odot}$ $<$ 5.2) O dwarfs, where the data of Martins et al. (\cite{vink:martins05}) 
fall well below predictions, by factors of 10 or more. The reason for this discrepancy is as yet not understood. 
Another area of discrepancies is that of the B supergiants where empirical rates have been found to be much lower 
than predicted rates (Vink et al. \cite{vink:vink00}, Trundle \& Lennon \cite{vink:trundle05}). 

The most worrisome however is the situation with garden-variety type O stars! 
Figure~\ref{vink:eta} shows that 
the Vink et al. (\cite{vink:vink00}) predictions are {\it lower} than the observed rates. The radio rates are likely to be 
upper limits as the radio free-free emission is a $\rho^2$ process, and any form of clumping leads to a maximal $\dot{M}$. 
Mokiem et al. (\cite{vink:mokiem07}) and Puls (this volume) noted that {\it if} this discrepancy is related to wind clumping -- and 
theoretical rates are unaffected -- the empirical $\rho^2$ mass-loss rates must 
be down-revised by a factor 2-3, suggesting a clumping factor $f$ $\sim$5. Recent massive star evolution models would not be 
effected by such modest clumping factors as these already use the theoretical Vink et al. (\cite{vink:vink00}) rates.

If $f$ $\sim$5 were universal, one would expect the empirical $\dot{M}$ for B supergiants
to be {\it lower} than indicated in Fig.~\ref{vink:eta} and the discrepancy would amount to an order of magnitude, or more. 
This implies there are some serious issues with our 
theoretical understanding of hot-star winds, and we need to reconsider even our most basic modelling assumptions, such as 
sphericity and homogeneity, which can be tested with linear polarimetry.

\section{Linear polarisation variability}
\label{vink:polvar}

\begin{figure}[H]
\begin{center}
\includegraphics
  [width=\columnwidth]{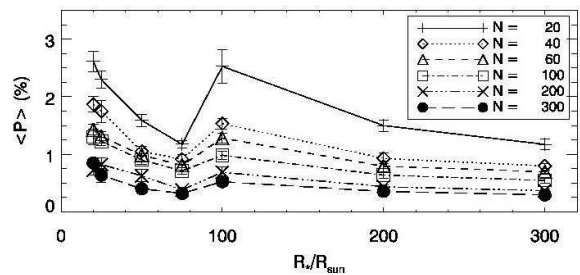}
\caption{Polarisation variability vs. stellar radius -- for different clump ejection rates.
$N$ $=$ $\dot{N}$ $t_{\rm fl}$, where  $\dot{N}$ is the clump ejection rate (related to the mass-loss rate) and 
$t_{\rm fl}$ $=$ $R/v_{\infty}$ (see Davies et al. \cite{vink:davies07}).
\label{vink:rstar}}
\end{center}
\end{figure}

Linear polarimetry is a tool to measure asymmetries. Davies et al. (\cite{vink:davies05}) performed 
a polarimetric survey of LBVs and found asymmetries in a majority of them. When the position angle (PA)
of the polarisation shows a straight line in the Stokes QU diagram this is generally attributed to a large-scale,
axi-symmetry, e.g. a disk. Davies et al. (\cite{vink:davies05}) found time-variable PAs for objects such as AG~Car 
and attributed these to wind clumping. Subsequently, Davies et al. (\cite{vink:davies07}) constructed an analytic clumping 
model, releasing clumps with a certain ejection rate per wind flow-time, $N$, from the wind base.
The average polarisation of the clump ensemble was calculated; the results for LBVs with a range of temperatures 
and radii are shown in Fig.~\ref{vink:rstar}. 

When the LBVs decrease their radii, the clumps become 
smaller and denser, and produce more polarisation. This behaviour reverses at the temperature of the 
bi-stability jump where the wind becomes faster, and the clumps spend less time at the wind base. As a result 
the polarisation drops. Figure~\ref{vink:rstar} also shows that when the ejection rate increases, the polarisation 
drops as the wind approaches that of a smooth outflow leading to zero polarisation.

\begin{figure}[H]
\begin{center}
\includegraphics
  [width=\columnwidth]{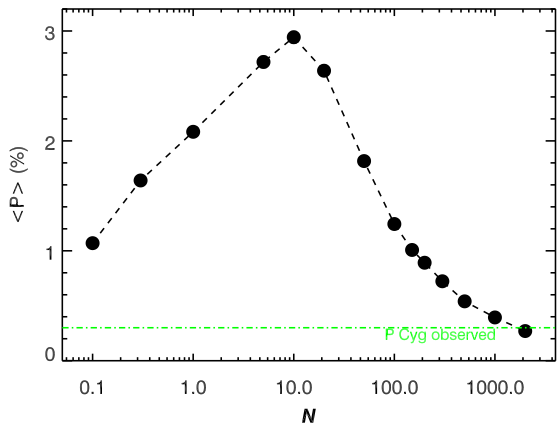}
\caption{Polarisation variability vs. clump ejection rate. The presence of a maximum 
implies there are two solutions compared to the observed level of P~Cyg.
\label{vink:pcyg}}
\end{center}
\end{figure}

We now consider the average polarisation as a function of ejection rate. 
The right-hand side of Fig.~\ref{vink:pcyg} shows the regime of many optically thin clumps. 
A maximum is reached at $N$ $\sim$10 where the clumps become optically thick and multiple scattering becomes 
important. 
The left-hand side represents the optically-thick clump regime. 
The observed level of polarisation of P~Cyg is shown as a horizontal line that intercepts both branches. 
The data either point to a wind with low $N$, or to one with $N$ $\sim$ 1000. 
We distinguish between these two branches using timescale information of the polarimetric variations. 
Preliminary results from our recent monitoring campaign indicate that the high-$N$ scenario is likely to be 
the correct one (Davies et al. {\it in prep}), which would suggest that LBV winds consist of thousands of 
optically thin clumps close to the photosphere.

\section{Conclusion}

Polarimetry is a tool to constrain the clumping properties of hot-star winds. 
This may become a powerful means by which to constrain non-LTE models and 
mass-loss predictions. 
We anticipate to witness an increased understanding of hot-star winds -- an important 
endeavour because of the impact mass loss has on massive star evolution modelling.

\end{multicols}

\end{contribution}

%%-------------------------------------------------------


\begin{thebibliography}{}

\bibitem[2007]{vink:benaglia07}
Benaglia, P., Vink, J.S., Marti, J., Maiz Apellaniz, J., Koribalski, B., Crowther, P.A., 2007, A\&A 467, 1265

\bibitem[2005]{vink:davies05}
Davies, B., Oudmaijer, R.D., Vink, J.S., 2005, A\&A 439, 1107

\bibitem[2007]{vink:davies07}
Davies, B., Vink, J.S., Oudmaijer, R.D., 2007, A\&A 469, 1045

\bibitem[1997]{vink:dekoter97}
de Koter, A., Heap, S.R., Hubeny, I., 1997, ApJ 477, 792

\bibitem[2007]{vink:galyam07}
Gal-Yam, A., Leonard, D.C., Fox, D.B., et al., 2007, ApJ 656, 372

\bibitem[2006]{vink:kotak06}
Kotak, R., Vink, J.S., 2006, A\&A 460, 5

\bibitem[2005]{vink:martins05}
Martins, F., Schaerer, D., Hillier, D.J., et al., 2005, A\&A 441, 735

\bibitem[2007]{vink:mokiem07}
Mokiem R., de Koter, A., Vink, J.S., Puls, J., et al., 2007, A\&A, in press

\bibitem[2007]{vink:pastorello07}
Pastorello, A., Smartt, S.J., Mattila,S., 2007, MNRAS 377, 1531

\bibitem[2007]{vink:smith07}
Smith, N., 2007, AJ 133, 1034

\bibitem[2005]{vink:trundle05}
Trundle, C., Lennon, D.J., 2005, A\&A 434, 677

\bibitem[1999]{vink:vink99}
Vink, J.S., de Koter, A., Lamers, H.J.G.L.M., 1999, A\&A 350, 181

\bibitem[2000]{vink:vink00}
Vink, J.S., de Koter, A., Lamers, H.J.G.L.M., 2000, A\&A 362, 295

\bibitem[2002]{vink:vink02}
Vink, J.S., de Koter, A., 2002, A\&A 393, 543

\end{thebibliography}
\end{document}